\documentclass[letter,scriptaddress,twocolumn, showkeys]{revtex4}
\usepackage{ amssymb }
	\usepackage{amsmath}
	\usepackage{makeidx}
	\usepackage{amsfonts}
	\usepackage[ansinew]{inputenc}
	\usepackage[usenames,dvipsnames]{pstricks}
	\usepackage{subfigure}
	\usepackage{epsfig}
	\usepackage{pst-grad} 
	\usepackage{pst-plot} 
	\usepackage[colorlinks,hyperindex]{hyperref}
	\hypersetup
	{
		colorlinks,%
		citecolor=blue,%
		linkcolor=blue,%
		urlcolor=blue,%
	}

	\newtheorem{defn}{Definition}

 	\newtheorem{assume}{Assumption}

 \newcommand{\CO}{\mathcal{O}}
 
\makeindex

\begin{document}

\title{Interface control and snow crystal growth}

\author{Jessica Li $^{a}$ and   Laura P. Schaposnik $^{b}$}

  \affiliation {(a) Kent Place School, 42 Norwood Avenue,  Summit, NJ 07902, USA \\
  (b) Department of Mathematics, University of Illinois, Urbana, IL 61801, USA} 

\begin{abstract}

The growth of snow crystals is dependent on the temperature and saturation of the environment.  In the case of dendrites,  Reiter's  local two-dimensional
model provides a realistic approach to the study of dendrite growth. In this paper 
we  obtain a new geometric rule that incorporates interface control, a basic mechanism of crystallization
that is not taken into account in the original Reiter's model. By defining two new variables, growth latency and growth direction, our improved model gives a realistic model not only for dendrite but also for plate forms.  \\

\end{abstract}

\maketitle
 

\section{Introduction}
\label{Introduction}

Snowflake growth is a specific example of crystallization - how crystals grow and create complex structures. Because
 crystallization corresponds to a basic phase transition in physics, and crystals make up the foundation of several major industries, studying snowflake growth helps gaining   understanding of how molecules condense to form crystals. This fundamental knowledge may help fabricate novel  types of crystalline materials \cite{[4]}. 
 
Whilst current computer modelling methods can generate snowflake images that successfully capture some basic features of real snowflakes, certain fundamental features of snowflakes growth are not well understood.   
One of the  key challenges has been that the snowflake growth models consist of a large set of PDEs, and as in many chaos theoretic problems, rigorous study is difficult. 

Snowflakes exhibit a rich combination of characteristic symmetry and complexity. The six fold symmetry is a result of the hexagonal structure of the ice crystal lattice, and the complexity comes from the random motion of individual snow crystals falling through the atmosphere:     
   
 \begin{figure}[h]
\centering
\resizebox{0.48\textwidth}{!}{%
  \includegraphics{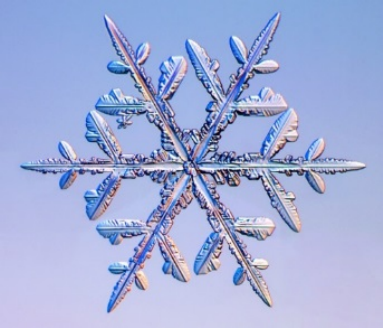}  \includegraphics{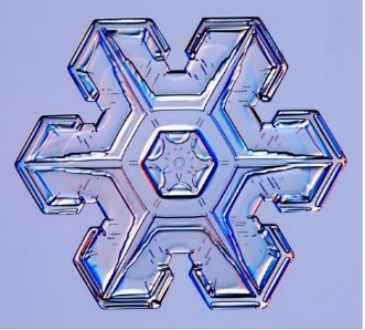}  \includegraphics{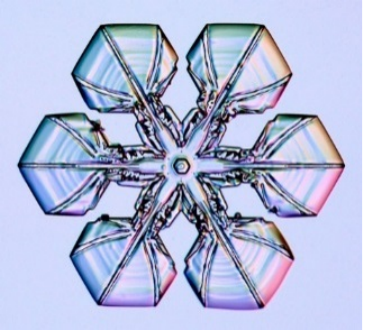}
  }
\caption{Examples of  real plate and dendrite snowflakes  \cite{[5]}. (a) Stellar dendrite (b) Stellar plate (c) Sectored plate.}\label{Figure 1}
\end{figure}
%
%
%
%

Scientific studies of snowflakes can be categorized into two main types. The first approach takes a macroscopic view by observing natural snowflakes in a variety of morphological environments characterized by temperature, pressure and vapour density (see  \cite{[6],[7],[8]} and references therein). The second type takes a microscopic view and investigates the basic  physical mechanisms governing the growth of snowflakes (e.g. see \cite{[4]}). While some aspects of snowflake growth (e.g., the crystal structure of ice) are well understood, many other aspects such as diffusion limited growth are at best understood at a qualitative level \cite{[4]}. 

Another approach in which snowflake growth is numerically simulated to produce images with mathematical models derived from   physical principles is through computer modelling (e.g. see  \cite{[2],[3],[9],[10],[11],[12]}). By comparing computer generated images with actual snowflakes, one can correlate the mathematical models and their parameters with physical conditions. While computer modelling can generate snowflake images that successfully capture some basic features of actual snowflakes, so far there has been only limited  analysis of these computer models in the literature. 

 In this paper we   analyze snowflake growth simulated by the computer models so as to connect the microscopic and macroscopic views and to further our understanding of snowflake physics.  The models that have been considered in the past are in essence chaos theoretic models, which is why they successfully capture the real world phenomena, but prove to be notoriously difficult to analyze rigorously. In this paper we study Reiter's popular model \cite{[11]}   using a combined approach of mathematical analysis and numerical simulation.

After reviewing Reiter's model in Section \ref{Reiter}, in Section \ref{General} we divide a snowflake image into main branches and side branches and define two new variables (growth latency and
growth direction) to characterize the growth patterns. In Section \ref{Growth2} we derive a closed form solution of the main branch growth latency using
a one dimensional linear model, and compare it with the simulation results using the hexagonal automata. Then, in Section \ref{Growthside} we discover a few
interesting patterns of the growth latency and direction of side branches. On the basis of the analysis and the principle of surface
free energy minimization, in Section \ref{enhanced}  we enhance Reiter's model and thus obtain realistic results both for dendrites and plate forms. We summarize our contributions and present a few future work directions in Section \ref{future}.


\section{An overview of Reiter's model}
\label{Reiter}

Reiter's model is a hexagonal automata which can be described as follows. Given a tessellation of the plane into   hexagonal cells, each cell $z$ has six nearest neighbours. We shall denote by $s_t (z)\in \mathbb{R}_{>0}$ the state variable of cell $z$ at time $t$ which gives the amount of water stored in  $z$. Then, cells are divided into three types:

\begin{defn} A cell $z$ is   {\rm frozen} if $s_t (z)\geq1$ (an F-cell) . If a cell is not frozen itself but at least one of the nearest neighbours is frozen, the cell is a {\rm boundary cell} (a B-cell). A cell that is neither frozen nor boundary is called {\rm nonreceptive} (an NR-cell). The union of frozen and boundary cells are called {\rm receptive cells} (R-cells).
 \begin{figure}[h]
\centering
\resizebox{0.5\textwidth}{!}{%
  \includegraphics{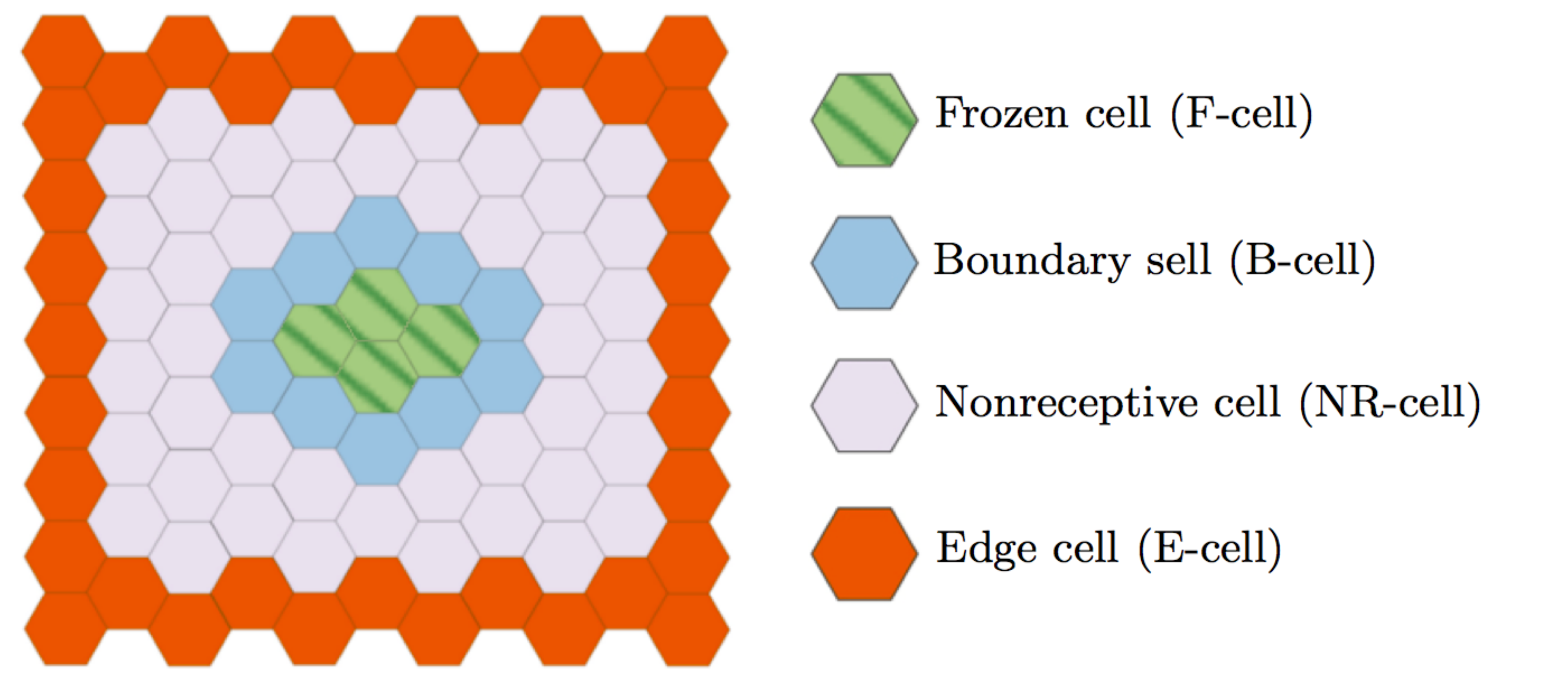}   
  }\caption{Classification of cells. 
  }
\end{figure}
\end{defn}

The initial condition in Reiter's model is 
\begin{eqnarray}
s_0(z)=\left\{\begin{array}
{ccc}
1&{\rm if}&z=\mathcal{O}\\
\beta&{\rm if}&z\neq\CO
\end{array}\right.  \nonumber
 \end{eqnarray}
where $\CO$ is the origin cell, 
and $\beta$ represents a fixed constant {\it background vapour level}. 
\begin{defn}Define the following functions on a cell $z$:
 the {\rm amount of water that participates in diffusion} $u_t (z)$; and the {\rm amount of water that does not participate} $v_t (z)$. Hence,  
 \begin{eqnarray} s_t(z)= u_t(z)+v_t(z) \end{eqnarray}
 and we let $v_t (z):=s_t (z)$ if $z$ is receptive, and $v_t (z):=0$ if $z$ is non-receptive.
\end{defn}
For $\gamma, \alpha$ two fixed constants representing {\it vapour addition} and {\it diffusion coefficients} respectively, 
in Reiter's model the state of a cell evolves as a function of the states of its nearest neighbours according to two local update rules that reflect the underlying mathematical models:
\begin{itemize}
\item 	{\it Constant addition}. For any receptive cell $z$, 
\begin{eqnarray}v_t^+ (z):=v_t^- (z)+\gamma	\label{(1)} \end{eqnarray}
\item 	{\it Diffusion}. For any cell $z$, 
\begin{eqnarray}u_t^+ (z):=u_t^- (z)+\frac{\alpha}{2} (\overline u^-_t (z)-u_t^- (z)),\label{(2)}\end{eqnarray}
\end{itemize}where we have used upper indices $^{\pm}$ to denote  new functions giving the state variable of a cell before and after a step is completed, and written $\overline{u}_t^- (z)$ for the average of $u_t^-$ for the six nearest neighbours of cell $z$.

The underlying physical principle of Eq.~(\ref{(2)}) is the diffusion equation \begin{eqnarray}\partial u/\partial t=a\nabla^2 u,\label{(3)}\end{eqnarray} where $a$ is a constant. Indeed, Eq.~(\ref{(2)}) is the discrete version of  Eq.~(\ref{(3)}) on the hexagonal lattice, and it states that a  cell $z$ retains $(1-\alpha/2)$  fraction of $u_t^- (z)$, uniformly distributes the remaining to its six neighbours, and receives $\alpha/12$ fraction from each neighbour. The total amount of $u_t (z)$ would be conserved within the entire system, except that a real world simulation consists of a finite number of contiguous cells. The cells at the edge of the simulation setup are referred to as {\it edge cells}, in which one sets $u_t^+ (z):=\beta$. Thus, water is added to the system via the edge cells in the diffusion process.
Combining the two intermediate variables, one obtains
\begin{eqnarray}s_{t+1} (z):=u_t^+ (z)+v_t^+ (z).\label{(4)}\end{eqnarray}
By varying the parameters $\alpha,\beta,\gamma$ in Reiter's model one can generate certain geometric forms of snowflakes observed in nature: 

 \begin{figure}[h]
\centering
\resizebox{0.4\textwidth}{!}{%
  \includegraphics{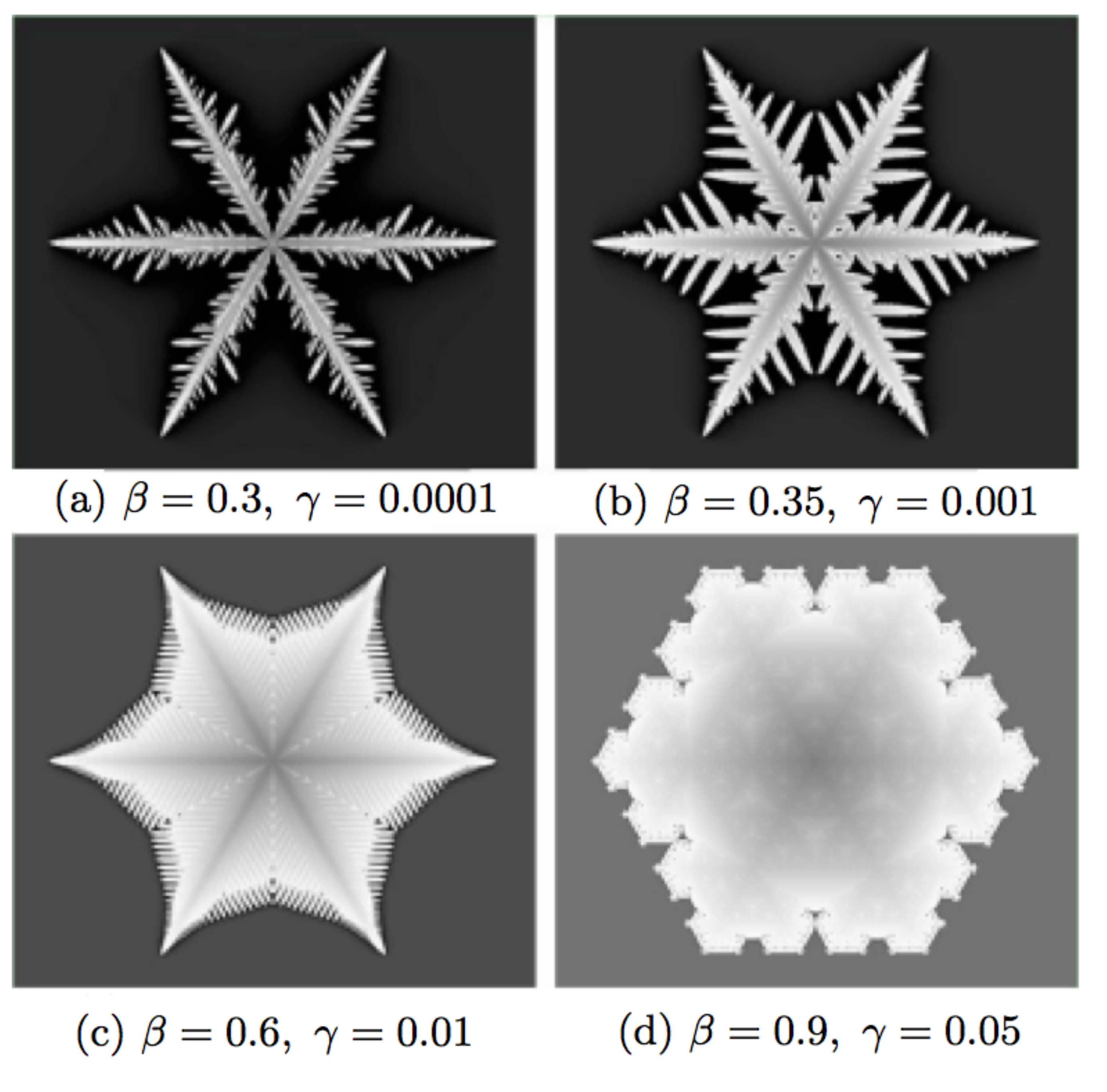}   
  }
\caption{Images generated by Reiter's model \cite{[11]} for $\alpha=1$.   } \label{Figure 2}
\end{figure}


\section{General geometric properties}
\label{General}

In what follows, we give new descriptions of snowflake growth and analyze them with a combined approach of mathematical analysis and numerical simulation by considering a coordinate system of cells as in Figure \ref{Figure 3}(a) below. A cell $z$ is represented by its coordinate $(i,j)$, for $i,j\in\mathbb{Z}$, with the origin $\CO=(0,0)$. Since there is a six fold symmetry, we only focus on one twelfth of the cells, marked as dark dots, for which $j\geq i\geq 0$: 
 \begin{figure}[h]
\centering
\resizebox{0.5\textwidth}{!}{%
  \includegraphics{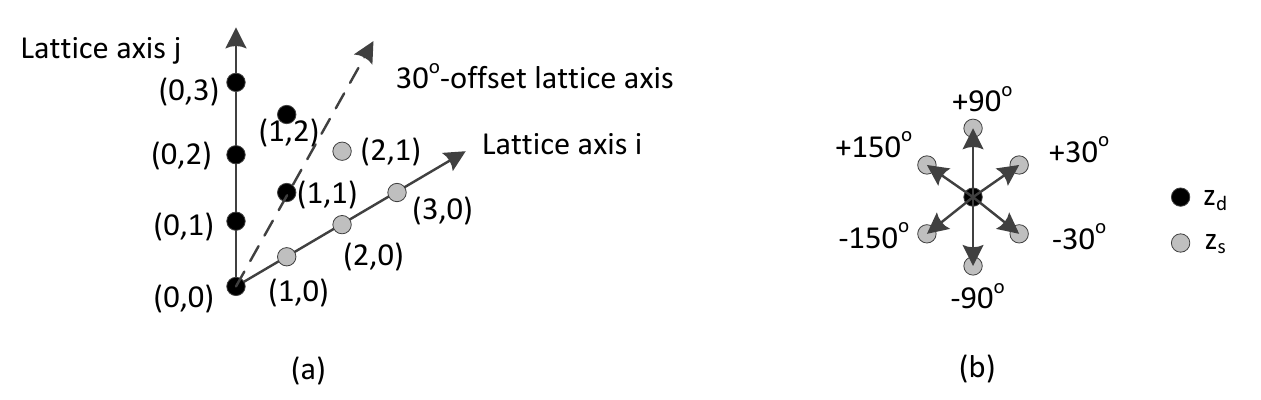}   
  }
\caption{(a) Coordinate system of hexagonal cells.\\ (b) Definition of growth directions in the coordinate system}\label{Figure 3}
\end{figure}

The images in Figure \ref{Figure 2} show that a crystal consists of six main branches that grow along the lattice axes, and numerous side branches that grow from the main branches in a seemingly random manner. The main and side branches exhibit a rich combination of characteristic symmetry and complexity which we shells rudy through the  {\rm  rate of water accumulation} of a cell $z$,    defined by $$\Delta s_t (z)=s_{t+1} (z)-s_t (z).$$  

At any time the set of all the R-cells is connected. Moreover since a frozen cell is surrounded by receptive cells and does not accumulate water via diffusion, and since water flows from nonreceptive cells to boundary cells, one has that the rate of water accumulation $\Delta s_t (z)$ of a cell satisfies   the following general geometric properties: 
\begin{enumerate}
\item[(A)]    For an NR-cell $z$, one has $0<s_t (z)\leq\beta$ and $\Delta s_t (z)\leq 0$. 
Suppose that an NR-cell $z$ is surrounded by R-cells and disconnected from the E-cells. If $\gamma>0$, there exists $t_0>0$ such that $s_{t_0 } (z)\geq1$ (i.e., a time $t_0$ in which the cell becomes frozen); otherwise,   the NR-cell will permanently remain non-receptive and never become frozen.

\item[(B)]  For a B-cell $z$ the quantity $ \Delta s_t (z)$ is the sum of  $\gamma$ and diffusion. If $\gamma>0$, there exists $t_0>0$ such that $s_{t_0} (z)\geq1$; otherwise, if cell $z$ is surrounded by a set of F-cells and disconnected from the E-cells, then as in (A), the cell will never become frozen:$$\lim_{t\rightarrow\infty}s_t (z)<1.$$
\item[(C)] For an F-cell $z$, one has $\Delta s_t (z)= \gamma$.

\item[(D)] The state variable of a cell is $s_t (z)=\beta$ only for  $z$ an E-cell.

 \end{enumerate}

Thus, for an NR-cell to become frozen, the cell goes through two stages of growth. First, the NR-cell loses vapour to other cells due to diffusion, as in   (A). Subsequently, it becomes a B-cell and accumulates water via diffusion and addition, as in   (B), until it becomes frozen and sees no benefit of diffusion, as in   (C). Becoming a B-cell is a critical event between the two stages.

 We focus on the second stage and define two new variables to characterize growth patterns.

\begin{defn}  The {\rm time to be frozen} of a cell $z$ is denoted by $T(z)$ and defined by the condition $s_{T(z)} (z)\geq 1$, and $s_t (z)<1$ for $t<T(z)$. Similarly, we define $B(z)$ as the {\rm first time to be boundary}. Finally, {\rm growth latency} is denoted by $L(z)$ and defined by $L(z):=T(z)-B(z)$.\label{B}\end{defn} 

A cell becomes a B-cell as one of its neighbouring cells has just become an F-cell, and thus it is useful to make the following definition in terms of redistribution of water: 

\begin{defn}Denote by $z_d$ a {\rm destination cell}, and by  $z_s$ a {\rm  source cell}. Then, the  
  {\rm growth direction} of cell $z_d$ is denoted by $g(z_d)$ and defined as the orientation of $z_s$ with respect to $z_d$, where the angle is with respect to the horizontal axis. The {\rm source-destination} cell relationship shall be denoted by $S(z_d):=z_s$. \label{defg}
\end{defn}
As shown in Figure \ref{Figure 3}(b), the angle is given relative to the horizontal direction in the coordinate system, and satisfies $$g(z_d)\in\{+30^\circ,-30^\circ,+90^\circ,-90^\circ,+150^\circ,-150^\circ\}.$$  Note that while the growth of $z_d$ is traced back to a unique $z_s$, a source cell may correspond to multiple destination cells.


\section{Growth of main branches}
\label{Growth2}

Consider cells  $(i,j)$  where $ i+j=K$ for a fixed $K$. These cells are all $K$ sites away from the origin $(0,0)$  on the grid. The main branch growth pattern is such that $ T(0,K)\leq T(i,j) $ and 
\begin{eqnarray} \begin{array}{rcclc}
T\left(\frac{K}{2},\frac{K}{2}\right)&\geq &T(i,j)& {\rm ~for~ even~ }&K,\\[5 pt]
T\left(\frac{K-1}{2},\frac{K-1}{2}\right)&\geq & T(i,j) &{\rm ~for~ odd~ }&K.\end{array}\nonumber
\end{eqnarray} 
Along the lattice $j$-axis, one has $g(0,j)=-90^\circ$ for all $j$. Hence, the snowflake growth is fastest along a lattice axis, which represents a main branch, and is the   slowest along the $30^\circ$-offset lattice axis.

We next develop a model to calculate the growth latency $L(0,j)$. As cell $(0,j) $  becomes frozen, cell  $ (0,j+1) $  becomes a boundary cell. Hence the first time to be boundary $B(0,j+1)=T(0,j)$, and thus one can calculate $T(0,j)$ as
\begin{eqnarray}
T(0,j)=T(0,0)+\sum_{k=1}^j L(0,k).
\end{eqnarray}

 In order to gain analytical understanding, we first study a one dimensional model. Consider a line of consecutive cells $z_0,z_1,\ldots,z_N$, where $Z_N$ is the edge cell. Initially cell $\CO$ is frozen. We focus on the growth period $[B(k),T(k)]$ in which cells $z_0,z_1,\ldots,z_{k-1}$  are frozen and cell $k$ grows from boundary to frozen. Since  Eq.~(\ref{(2)}) describes the diffusion dynamics of vapour being transferred from the edge cell to cell $z_k$,  and cell $z_k$ accumulates water via addition  Eq.~(\ref{(1)}), to derive an analytical solution, we make the following assumption which we justify shortly.

\begin{assume} For $t\in[B(z_k),T(z_k)]$, assume that in  Eq.~(\ref{(2)}) one has $$u_t^+ (z_i)=u_t^- (z_i),$$ for $k+1\leq i\leq N$,  for $N$ as above and $B(z_k),T(z_k)$ as in Definition \ref{B}. Therefore, the vapour distribution reaches a {\rm steady state}, denoted as $\mu(i|k)$.\label{Assumption 4.1}
\end{assume}

From Assumption \ref{Assumption 4.1}, we can ignore the notations of $\pm$, and reduce  Eq.~(\ref{(2)}) to the linear equation
\begin{eqnarray}
\mu(i|k)=\frac{1}{2} \left(\begin{array}{c}~\\~\end{array} \hspace{-0.1in}\mu(i-1|k)+\mu(i+1|k)\right).\label{(6)}
\end{eqnarray}
Moreover, with the boundary conditions $\mu(k|k)=0$ and $\mu(N|k)=\beta$, the vapour distribution can be written in a closed form as follows:
\begin{eqnarray}
\mu(i|k)=\frac{i-k}{N-k} \beta,  ~ {\rm for}~i=k,\ldots,N,\label{(7)}
\end{eqnarray}
which graphically represents a line that connects the two boundary condition points. 

We shall now explain why   Assumption \ref{Assumption 4.1} is well motivated. Suppose that the steady state distribution  Eq.~(\ref{(7)}) is already reached at $t=B(z_k)$, this is, 
\[s_{B(z_k)} (z_i)=\mu(i|k-1)=\frac{ i-(k-1)}{N-(k-1) } \beta~ {~\rm for~} i=k,\ldots,N.\]  
Then,  $s_t (z_i)$ evolves in the interval of $(B(z_k),T(z_k)]$ in the following manner. For $N\gg k$, one has
 \[s_{B(z_k)} (z_k)=\mu(k|k-1)=\frac{1}{(N-k+1)} \beta\approx 0.\] 
Thus it is reasonable to assume $$L(z_k)=T(z_k)-B(z_k)\gg 1$$ because cell $z_k$ will take several simulation steps to reach $s_{T(z_k)} (z_k)\geq1$. Moreover, 
\[|s_{B(z_k)} (z_i)-\mu(i|k)|=\left|\frac{i-(k-1)}{N-(k-1) } \beta- \frac{(i-k)}{(N-k)} \beta\right|\ll 1\] 
for $N\gg k$. Thus, in each simulation 
step for time $t\in(B(z_k),T(z_k)]$, the function $s_t (z_i)$ only varies slightly and can be considered approximately constant. Hence, $u_t^+ (z_i)=u_t^- (z_i)$.

From  Eq.~(\ref{(2)}) and  Eq.~(\ref{(6)}), we may estimate $u_t^+ (z_k)$ by $$\hat u_t^+ (z_k):=\frac{\alpha}{4}\frac{  1}{N-k} \beta.$$ Moreover, since $u_t^- (z_k)=0$ it follows that we can further estimate $\Delta  s_{t} (z_k)$ and $L(z_k)$ by
\begin{eqnarray}
\Delta \hat s_{t} (z_k)&:=&\frac{\alpha}{4} \frac{1}{N-k} ~\beta+\gamma,\label{(8)}\\
\hat L(z_k)&:=&\frac{1-s_{B(z_k)} (z_k)}{\Delta \hat s_t (z_k) }  = \frac{1-\frac{1}{N-k+1} \beta}{\frac{\alpha}{4}  \frac{1}{N-k}~ \beta+\gamma}.\label{(9)}
\end{eqnarray}

\begin{figure}[h]
\centering
\resizebox{0.4\textwidth}{!}{%
  \includegraphics{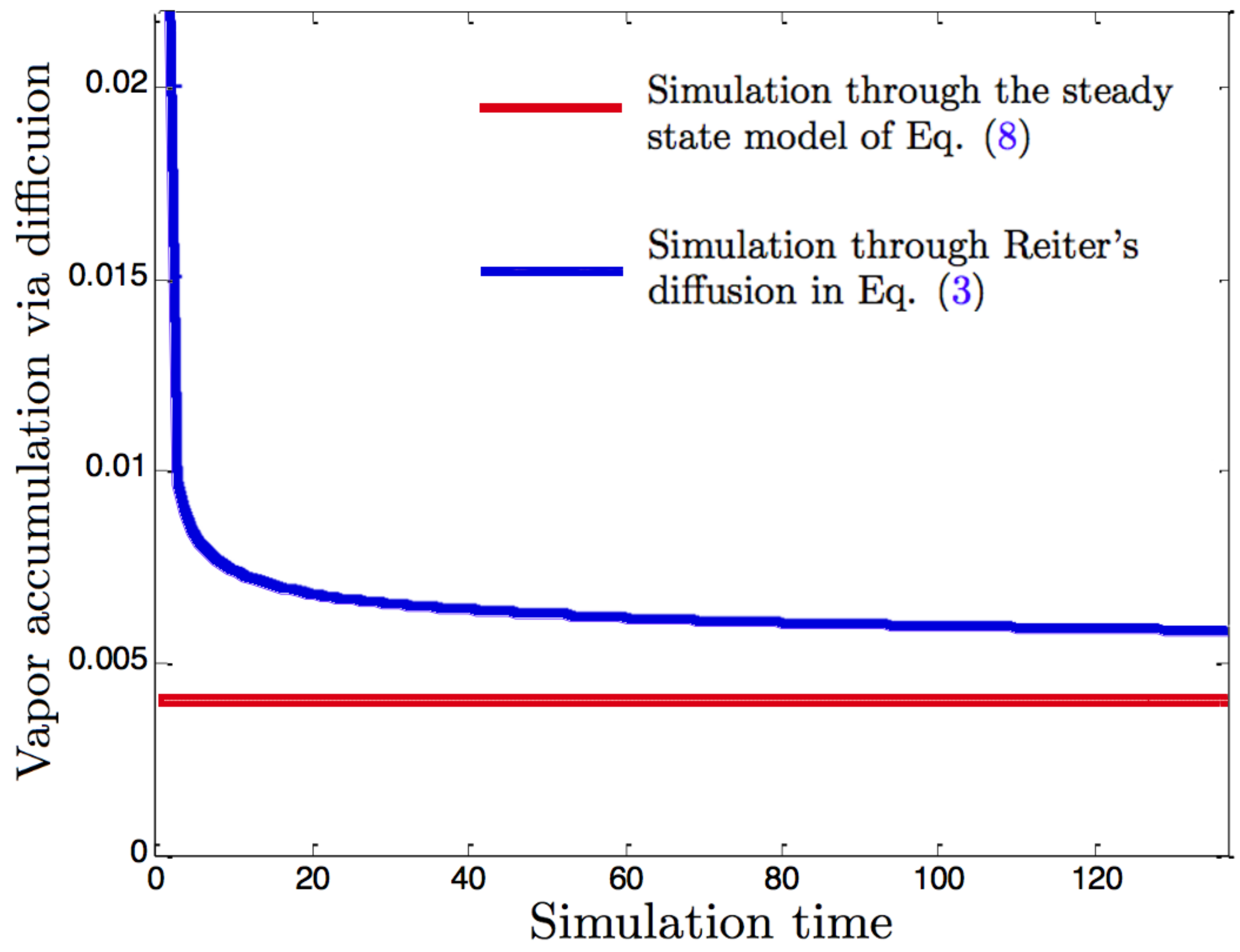}        
  }
\caption{Comparison of vapour accumulation simulations for the parameters 
$\alpha=1, ~\beta=0.4, ~\gamma=0.001$. }\label{Figure 4a}
\end{figure}

In the one dimensional model with $N=50$, we may compare the vapour accumulation in every simulation step, as the simulation proceeds from the time when cell  $k=25 $ just becomes boundary to the time when it becomes frozen. Figure \ref{Figure 4a}   compares $\Delta s_t (z_k)$  at cell $z_k$ determined by the simulation, and $\Delta \hat s_t (z_k) $  predicted by Eq.~(\ref{(8)}) for time  $t\in[B(z_k),T(z_k)]$.  Initially $s_t (k)\ll\mu(i|k)$, and $\Delta s_t (k)\gg \Delta \hat s_t (k)$. After about 5 simulation steps, $\Delta s_t (k)$ drops to a flat plateau, which is approximately equal to $s_t (k)$. At any time $t$, one observes that $\Delta \hat s_t (k)\leq \Delta s_t (k)$.

\newpage

\begin{figure}[h]
\centering
\resizebox{0.4\textwidth}{!}{%
   \includegraphics{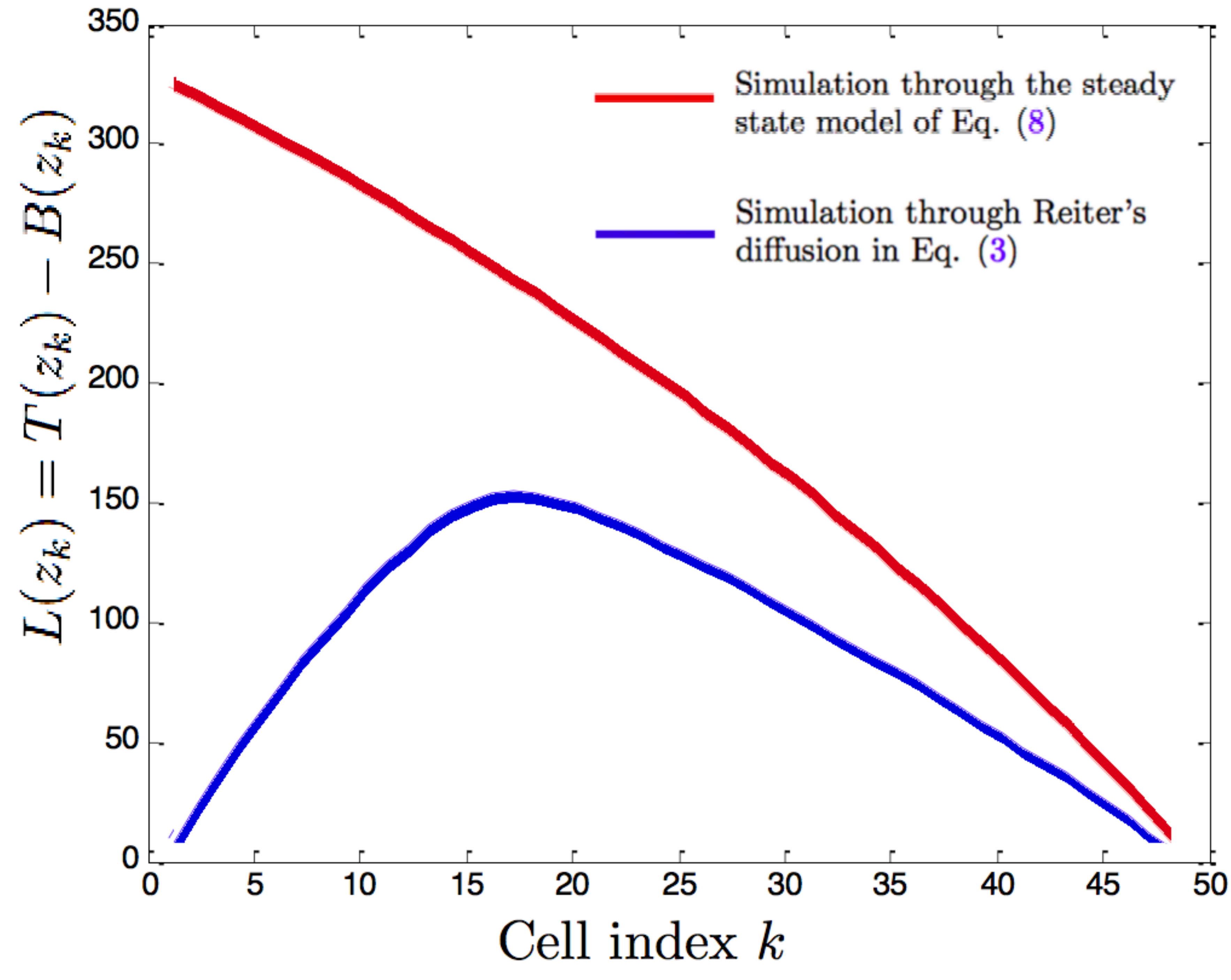}   
  }
\caption{ 
Comparison of growth latency simulations for the parameters 
$\alpha=1, ~\beta=0.4, ~\gamma=0.001$. }\label{Figure 4b}
\end{figure}

One may also model  $L(z_k)=T(z_k)-B(z_k)$  as a function of cell index of the cells.  In the one dimensional model with $N=50$,  Figure \ref{Figure 4b} compares $L(k)$  determined by the simulation, and $\hat L(z_k)$ predicted by  Eq.~(\ref{(9)}) as the snowflake grows from the origin $\CO$ to the edge cell. For any $k$, one observes that $L(z_k)<\hat L(z_k)$. 
This phenomenon is expected, since  by solving the above PDE's one has that
there exists $\alpha>0$ such that at any time instance $t\in[B(z_k),T(z_k)]$, for $ i=k,\ldots,N, $ one has \[\mu(i|k)\leq s_t (z_i) ~{~\rm and~} ~\Delta \hat s_t (z_k)\leq \Delta s_t (z_k).\] As a result, $\hat L(z_k)\geq L(z_k)$.
 
Equation Eq.~(\ref{(9)}) predicts that $\hat L(z_k)$ drops monotonically with $ k$. In simulation, we observe that in the beginning the cells grow from boundary to frozen very quickly, well before the steady state is reached. As a result, the steady state Assumption \ref{Assumption 4.1} does not hold in that time period.          Figure  \ref{Figure 4b} shows that $L(z_k)$ first increases, then drops, and eventually matches the prediction $\hat L(z_k)$.

Finally, we return to the two dimensional hexagonal cellular case. With a similar steady state assumption, we can reduce the PDE to a set of linear equations similar to  Eq.~(\ref{(6)}). However, the geometric structure is much more complex than the one dimensional case. As a result, it is difficult to derive a closed form formula of the vapour distribution similar to Eq.~(\ref{(7)}).   

\begin{figure}[h]
\centering
\resizebox{0.4\textwidth}{!}{%
  \includegraphics{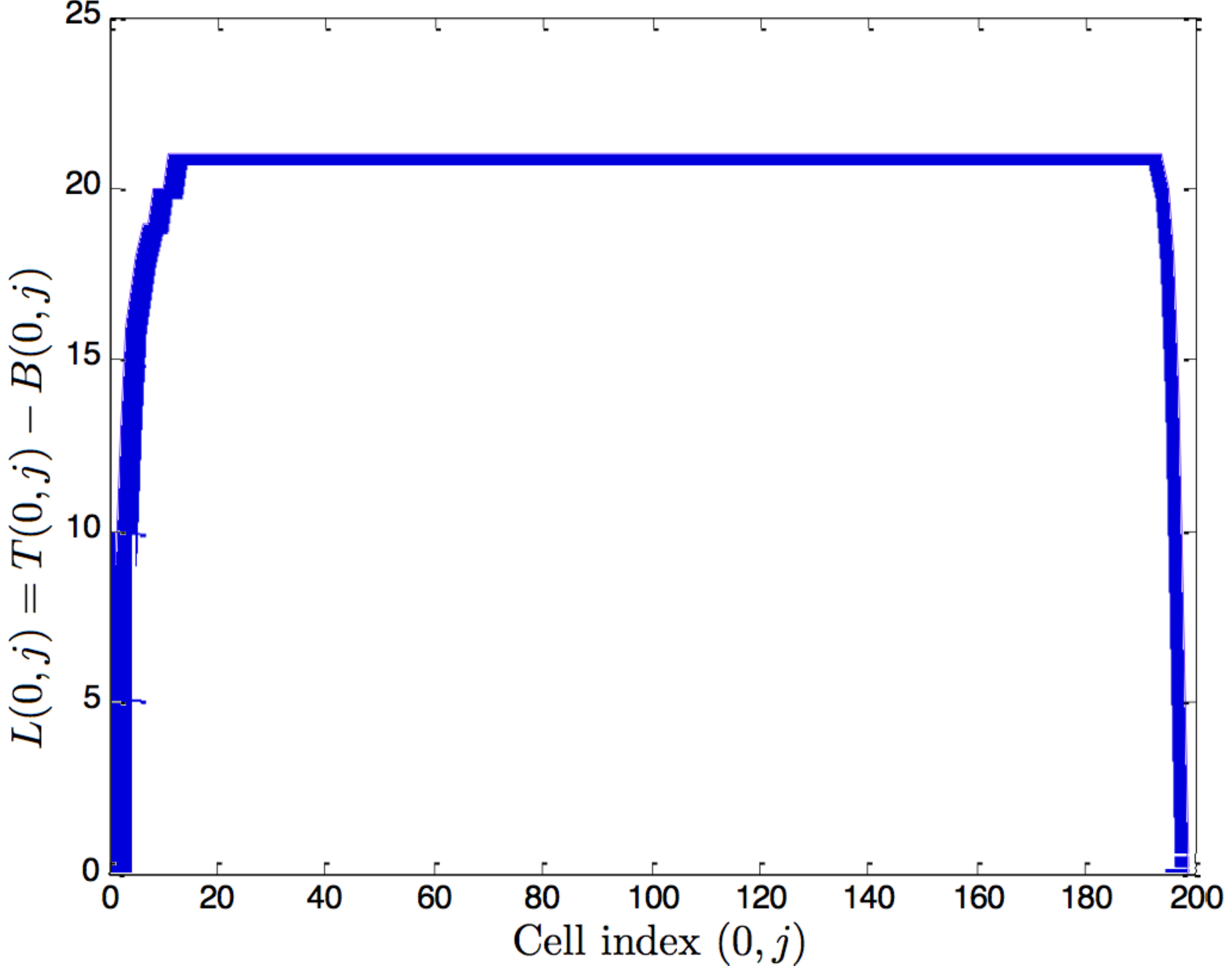}      }\caption{ $T(0,j)-B(0,j)$ of cells $ (0,j)$ along a main branch for $j=1,2,\ldots,$ in the two dimensional scenario. Here cell $(0,200)$  is an edge cell,  and  $\alpha=1$, $\beta=0.4$, $\gamma=0.001$. }\label{Figure 5}
\end{figure}

Figure \ref{Figure 5} below plots $L(0,j)$ along a main branch. Comparison with Figure \ref{Figure 4b} indicates a similarity between the one dimensional and two dimensional cases in that $L(z)$ increases as the snowflake grows from the origin. However, in the two dimensional case, we observe from simulations that $L(0,10)=L(0,11)=\ldots=L(0,195)$.  When the snowflake grows close to the edge cell, it experiences some edge effect in the simulation where $L$ drops drastically. This indicates that somewhat surprisingly $\Delta s_t (0,j)$ remains almost constant as the snowflake grows along the main branch.

\section{Growth of side branches}
\label{Growthside}

While the main branches of snowflakes represent clean six fold symmetry, the side branches exhibit characteristic features of chaotic dynamics: complexity and unpredictability. Reiter's model is completely deterministic with no noise or randomness involved, and yet the resultant snowflake images are sensitive to the parameters $\alpha,\beta$, and $\gamma$ in a chaotic manner. Chaos may appear to be the antithesis of symmetry and structure. Our goal in this section is to discover growth patterns that emerge from seemingly chaotic dynamics. 

\begin{defn}\label{Definitions 5.1}
 Starting from a cell $z_0$ on the j-axis main branch, the set of consecutive frozen cells in the i-axis direction are referred to as {\rm side branch} from cell $z_0$. We shall denote  by $z_E(z_0 ) $  the {\rm outmost cell} or {\rm tip},   by $E(z_0 )$ the {\rm length of the side branch}, and the side branch itself by $\Phi(z_0 ):=\{z_0,\ldots,z_E(z_0 )  \}$. 
\end{defn}

\begin{figure}[h]
\centering
\resizebox{0.5\textwidth}{!}{%
  \includegraphics{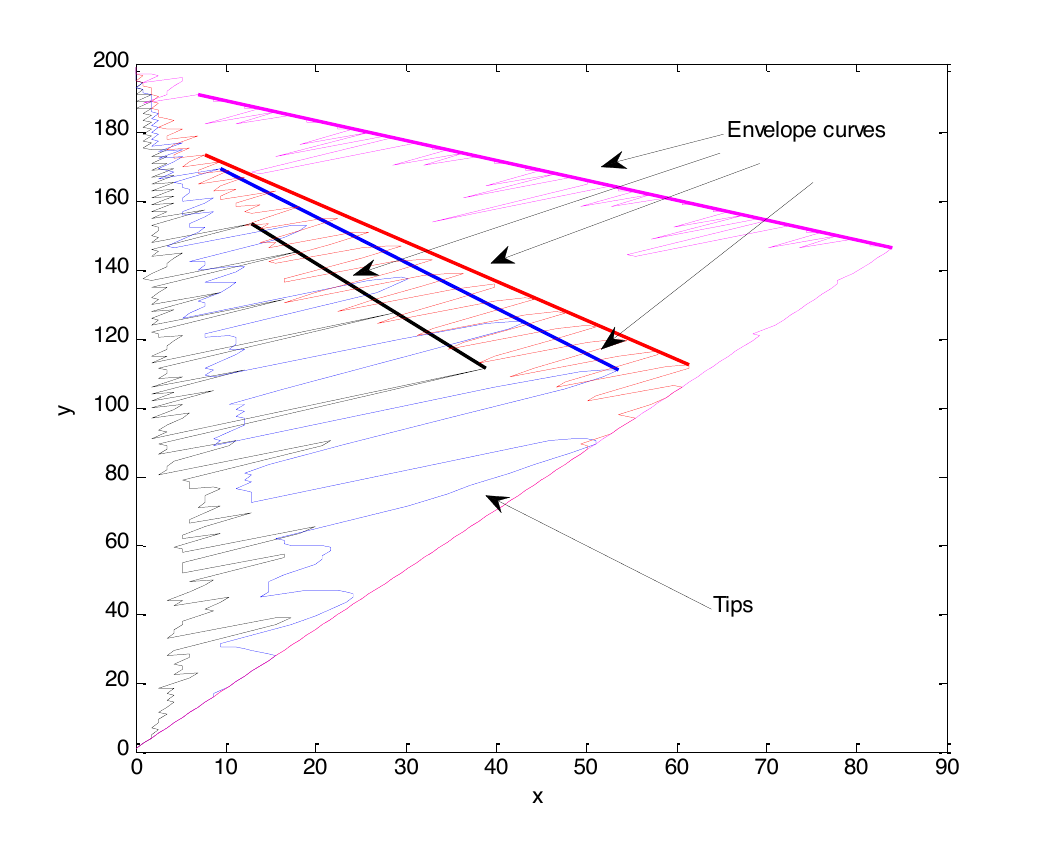}      }
\caption{
Plots of the tips (thin curves) and envelope curves (thick curves) of the side branches from the $j$-axis main branch using the parameters of the four example images in Figure \ref{Figure 2}. Due to symmetry, we focus on one set of side branches that grow from the right side of lattice $j$-axis. Here, the black curve represents Figure \ref{Figure 2}(a), blue for Figure \ref{Figure 2}(b), red for Figure \ref{Figure 2}(c), and magenta for Figure \ref{Figure 2}(d). The  axes here are the horizontal and vertical axes of the coordinate system.} \label{Figure 6}
\end{figure}

In what follows, we study the growth latency of side branches. Figure \ref{Figure 6} below plots the tips of the side branches that grow from the j-axis main branch using the parameters of the four images in Figure \ref{Figure 2}.

 Due to the chaotic dynamics, the lengths of the side branches vary drastically with $z_0$ in a seemingly random manner. For image (a), most of the side branches are short and only a small number stand out. The opposite holds for image (d). The scenarios are in between for images (b) and (c). The length of the side branches is indicative of the growth latency. The long side branches represent the ones that grow fastest. In Figure \ref{Figure 6} we connect the tips of the long side branches to form an {\it envelope curve} that represents the frontier of the side branch growth. The most interesting observation is that the envelope curve can be closely approximated by a straight line for the most part. Recall that the growth latency of the main branch is a constant. Thus we infer that the growth latency of the long side branches is also constant. Denoting by $L_M$ and $L_S$  the {\it growth latencies } of the main and long side branches respectively, one has that
\begin{eqnarray}
\frac{L_M}{L_S} = \frac{\sin 2\pi / 3 - \theta}{\sin \theta} 
\end{eqnarray}
where $\theta$ is the angle between the envelope curve straight line and the $j$-axis. As a specific example, for the magenta curve, the envelope curve of the long side branches grows almost as fast as the main branch, such that $\theta\approx \pi/3$ and the resultant image, appearing in  Figure \ref{Figure 2}(d),  is roughly a hexagon.

\begin{figure}[h]
\centering
\resizebox{0.45\textwidth}{!}{%
  \includegraphics{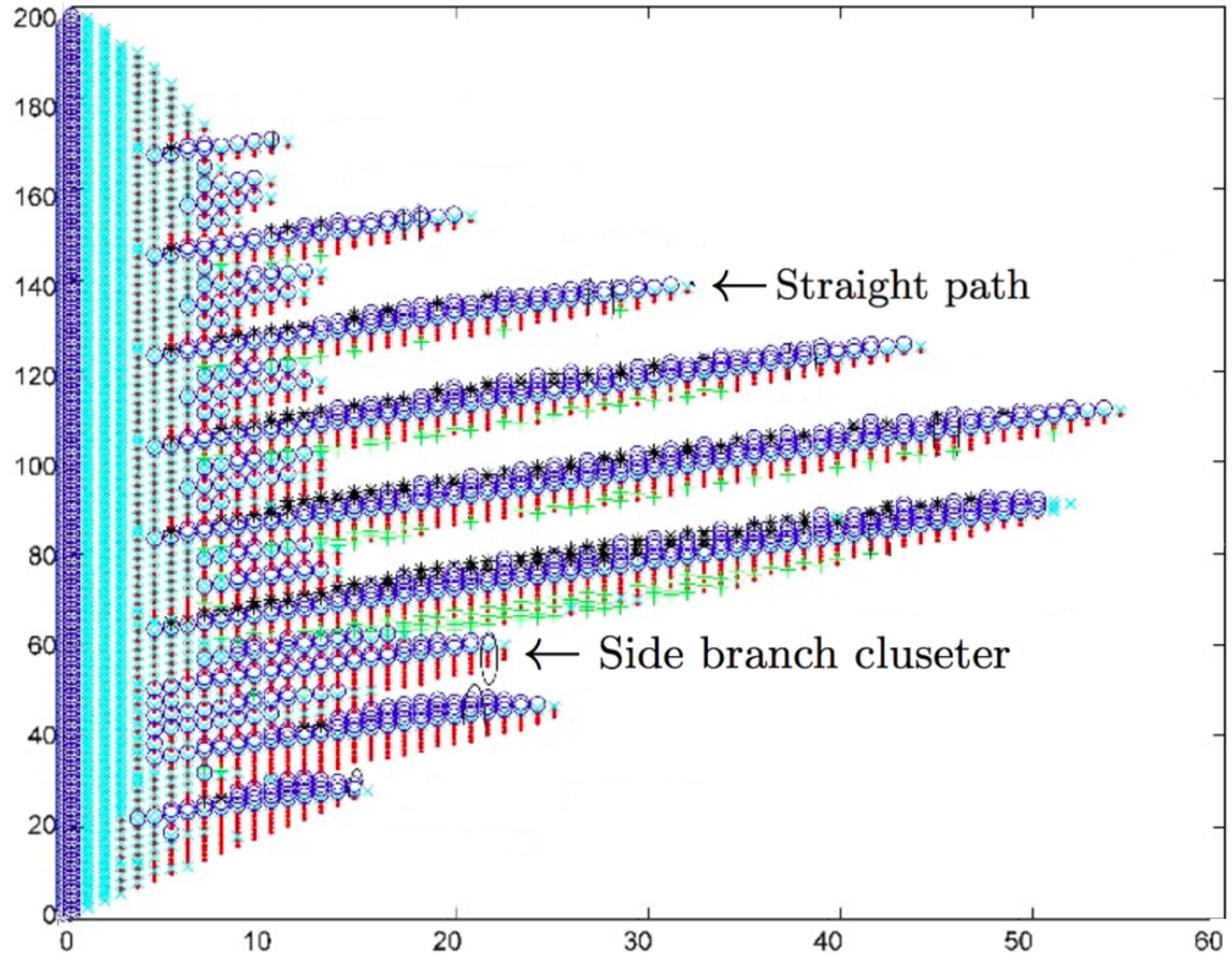}  }
\caption{ 
Trace of relative orientations of source cells with respective destination cells. A destination cell becomes boundary because a source cell, which is one of the neighbours of the destination cell, becomes frozen. Legend is as follows: $magenta ~\blacksquare:+30^\circ, ~~black~ \star:-30^\circ, ~~green ~+ :+90^\circ, \\
blue ~\circ :-90^\circ,~~ red ~\cdot:+150^\circ,~~ cyan~\times :-150^\circ,$ where the parameters are  $\alpha=1,~\beta=0.35,~\gamma=0.001$. Note that not all straight paths are labeled. The axis are as in Figure  \ref{Figure 3}(a). } \label{Figure 7}
\end{figure}

We shall consider next the growth directions of the cells on side branches. Figure \ref{Figure 7} below plots the trace of the growth direction $g(z)$ (see Definition \ref{defg}) as a snowflake develops in the simulation. The corresponding snowflake image is shown in Figure \ref{Figure 2}(b). When a cell $z$ becomes boundary, we mark the cell to indicate $g(z)$ using the legend labeled in the figure. If a cell never becomes boundary, no mark is made. All side branches grow from the $j$-axis main branch, starting in the direction parallel to the $i$-axis. Subsequently, a side branch may split into multiple directions. Indeed, all six orientations have been observed and the dynamics appear chaotic as $g(z)$  appears unpredictable. However, we do find an interesting pattern described below.

\begin{defn}\label{Definitions 5.2}
A {\rm straight path} from a cell $z_0$ on the $j$-axis main branch is the set of consecutive frozen cells in the $i$-axis direction satisfying $z_{i-1}=S(z_i )$. The number of consecutive cells satisfying $z_{i-1}=S(z_i )$ is the {\rm length} $F(z_0)$. We then denote a {\rm straight path} from a cell $z_0$  by $\Psi(z_0 ):=\{z_0,z_1,z_2,\ldots,z_{F(z_0 )}  \}$.  \end{defn}

Comparison between Definition \ref{Definitions 5.1} and Definition \ref{Definitions 5.2} shows that the paths are nested, i.e. $\Psi(z_0 )\subset \Phi(z_0 )$, and hence the lengths satisfy $F(z_0 )\leq E(z_0 ).$ When a cell $z_{i-1}$ on the straight path becomes frozen, it triggers not only $z_i$  in the i-axis direction but also other neighbours to become boundary, resulting in growth in other directions, which we call {\it deviating paths}. The straight and deviating paths collectively form a {\it side branch cluster}.

\begin{defn}
 The set of frozen cells that can be traced back to a cell on the straight path from cell $z_0$ on the j-axis main branch is referred to as a {\rm side branch cluster} and  denoted by $\Theta(z_0 )$.
\end{defn}

\begin{figure}[h]
\centering
\resizebox{0.45\textwidth}{!}{%
  \includegraphics{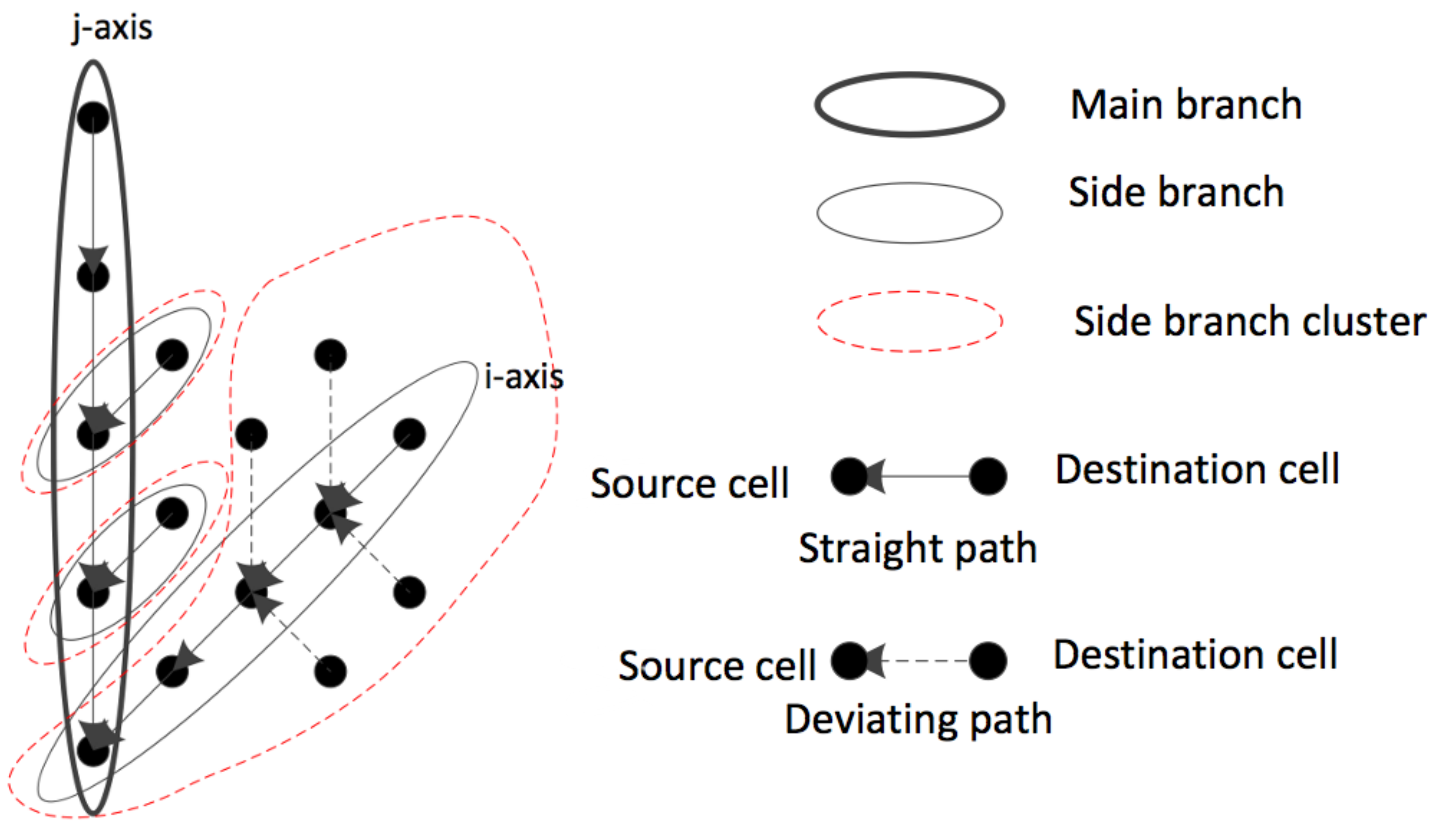}  }
\caption{An arrow linking two cells indicates the source/destination relationship. }\label{Figure 8}
\end{figure}

A side branch cluster is a visual notion of a collection of side branches that appear to grow together. Figure \ref{Figure 7} shows several side branch clusters and the cells on the corresponding straight path marked with cyan $\times$. Compared with the straight paths, the deviating paths do not grow very far, because they compete with other straight or deviating paths for vapour accumulation in diffusion. On the other hand, the competition with the deviating paths slows down or may even block the growth of a straight path. When a straight path is blocked, the straight path is a strict subset of the corresponding side branch. This scenario is illustrated in Figure \ref{Figure 8}, where three side branches are shown. 

The straight path of the middle side branch is blocked by a deviating path of the lower side branch, which grows into a sizeable side branch cluster. Through the above definitions one has  that if  there exists a cell $z$ such that $z\in\Theta(z_0)$ and $z\not\in \Phi(z'_0 )$, then the paths are nested $\Psi(z'_0 )\subset\Phi(z_0')$.
Moreover,  the straight path determines the length of the side branch cluster: 

\begin{defn}
Denote by $D(z,z_0 )$ the {\rm distance between $z_0, z\in \Theta(z_0 )$}, defined as the smallest number of sites on the lattice between $z$ and $z_0$. The {\rm length} of $\Theta(z_0 )$ is \[D(z_0 ):=\max_{z~ \in ~\Theta (z_0 )}D(z,z_0 ).\] 
\end{defn}

Through the above definition one can show that there are $K$ cells $z_i\in \Theta(z_0 )$ such that the distances satisfy $D(z_i,z_0 )=D(z_0 )$ for $i=1,\ldots,K$ with $K\geq 2$. Furthermore, there exists  $z_i\in \Psi(z_0 )$ for $1\leq i\leq K$, and thus  $D(z_0 )=F(z_0 )$,  the length of $\Psi(z_0)$ as in Definition \ref{Definitions 5.2}.
 

\section{An enhanced Reiter's model}
\label{enhanced}

Plates and dendrites are two basic types of regular, symmetrical snowflakes. We observe that while the dendrite images in Figure \ref{Figure 2}(a)(b) generated by Reiter's model resemble quite accurately the real snowflake in Figure \ref{Figure 1}(a), as seen in Figure \ref{Figure 2}(c)(d) and Figure \ref{Figure 1}(b)(c), the plate images differ significantly. The plate images in Figure \ref{Figure 2}(c)(d) is in effect generated as a very leafy dendrite. One of the reasons that Reiter's model is unable to generate plate images realistically is that the model only includes diffusion, thus not taking into account the effect of local geometry. 

As described in \cite{[1]}, two basic types of mechanisms contribute to the solidification process of snowflakes: diffusion control and interface control. Diffusion control is a non geometric growth model, where snowflake surfaces are everywhere rough due to diffusion instability, a characteristic result of chaotic dynamics. For example, if a plane snowflake surface develops a small bump, it will have more exposure into the surrounding vapour and grow faster than its immediate neighbourhood due to diffusion. Interface control is a geometric mechanism where snowflake growth only depends on local geometry, i.e., curvature related forces. In the small bump example, the surface molecules on the bump with positive curvature have fewer nearest neighbours than do those on a plane surface and are thus more likely to be removed, making the bump move back to the plane. Interface control makes snowflake surfaces smooth and stable, and it is illustrated in Figure \ref{Figure 9} below. 

 \begin{figure}[h]
\centering
\resizebox{0.3\textwidth}{!}{%
  \includegraphics{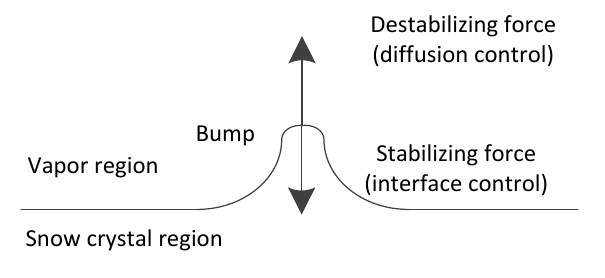}  }
\caption{ Two competing forces of diffusion control and interface control that determine snowflake growth. } \label{Figure 9}
\end{figure}

In summary, snowflake growth is determined by the competition of the destabilizing force (diffusion control) and stabilizing force (interface control). In the absence of interface control, Reiter's model is unable to simulate certain features of snowflake growth. 

\begin{figure}[h]
\centering
\resizebox{0.3\textwidth}{!}{%
  \includegraphics{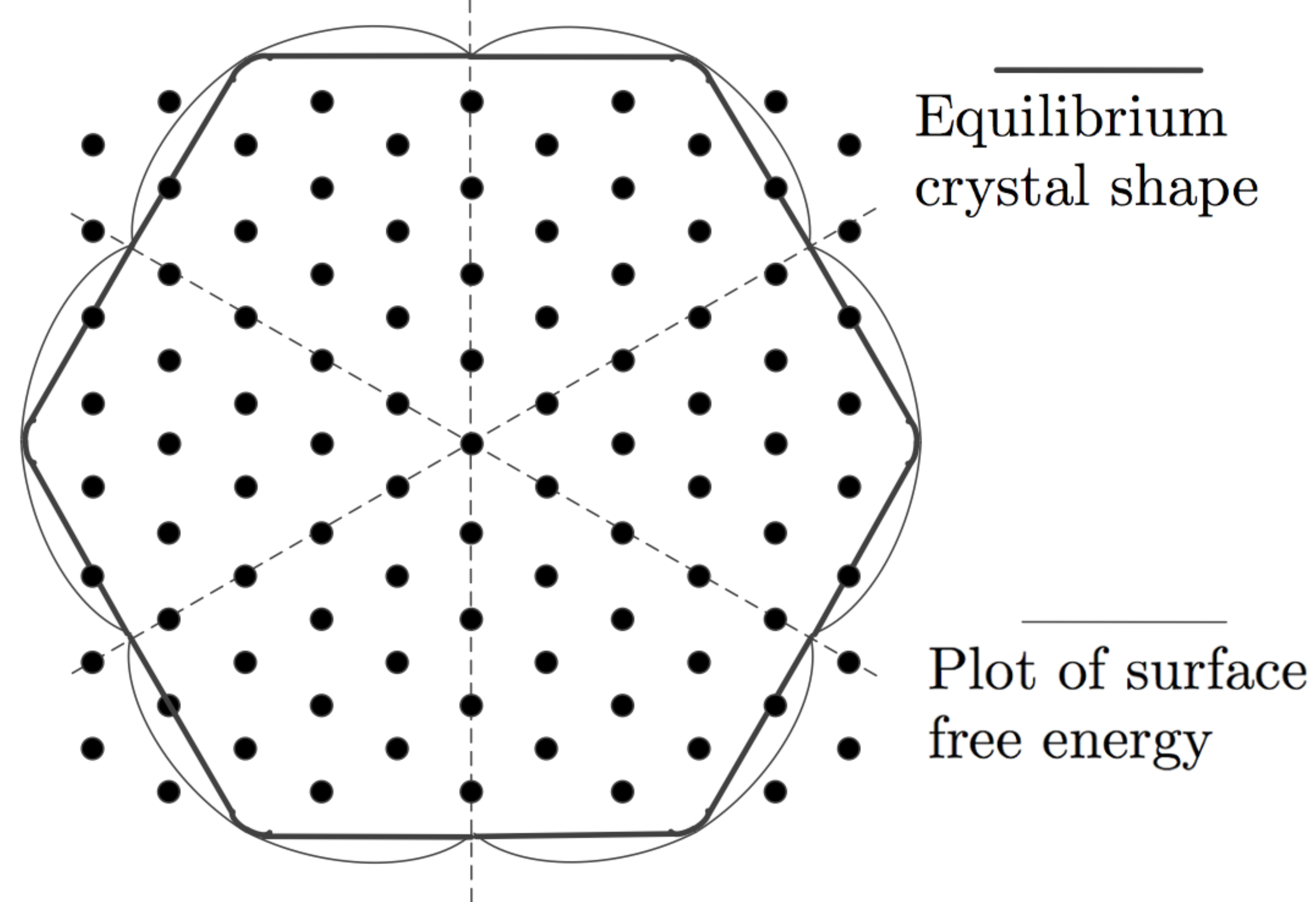} }
\caption{ 
Surface free energy of snowflake as a function of direction and equilibrium crystal shape of snowflake derived from surface free energy plot with Wulff construction \cite{[1]}.   }\label{Figure 10a}
\end{figure}

The interface between the snowflake and vapour regions has potential energy, called surface free energy, due to the unfilled electron orbitals of the surface molecules. The {\it surface free energy} $\gamma(n)$  as a function of direction $n$, is determined by the internal structure of the snowflake, and in the case of a lattice plane, is proportional to lattice spacing in a given direction. Figure \ref{Figure 10a} below plots the surface free energy $\gamma(n)$  of a snowflake as a function of the direction $n$.

The equilibrium shape of the interface is the one that minimizes the total surface free energy for a given enclosed volume. Wulff construction (see \cite{[1]}) can be used to derive the equilibrium crystal shape $\mathcal{W}_\gamma$  from the surface free energy plot $\gamma(n)$:
\begin{eqnarray}
\mathcal{W}_\gamma:=\{r~|~r\cdot n\leq \gamma(n),~\forall n \}.
\end{eqnarray}

Wulff construction states that the distances of the equilibrium crystal shape from the origin are proportional to their surface free energies per unit area. Figure \ref{Figure 10a} plots the equilibrium crystal shape of snowflake. Moreover, it shows that due to interface control, snowflake growth is the slowest along the lattice axes, and the fastest along the 30°-offset lattice axes. 

 This can be explained intuitively. Snowflake grows by adding layers of molecules to the existing surfaces. The larger the spaces between parallel lattice planes, the faster the growth is in that direction. This effect is completely opposite to the diffusion control we have studied in  Section 4, where snowflake grows fastest along the lattice axes. This is an example of competition between diffusion control and interface control.
 
 We next propose a new geometric rule to incorporate interface control in Reiter's model. The idea is that the surface free energy minimization forces the lattice points on an equilibrium crystal shape to possess the same amount of vapour so that the surface tends to converge to the equilibrium crystal shape as the snowflake grows.
 
\begin{figure}[h]
\centering
\resizebox{0.3\textwidth}{!}{%
 \includegraphics{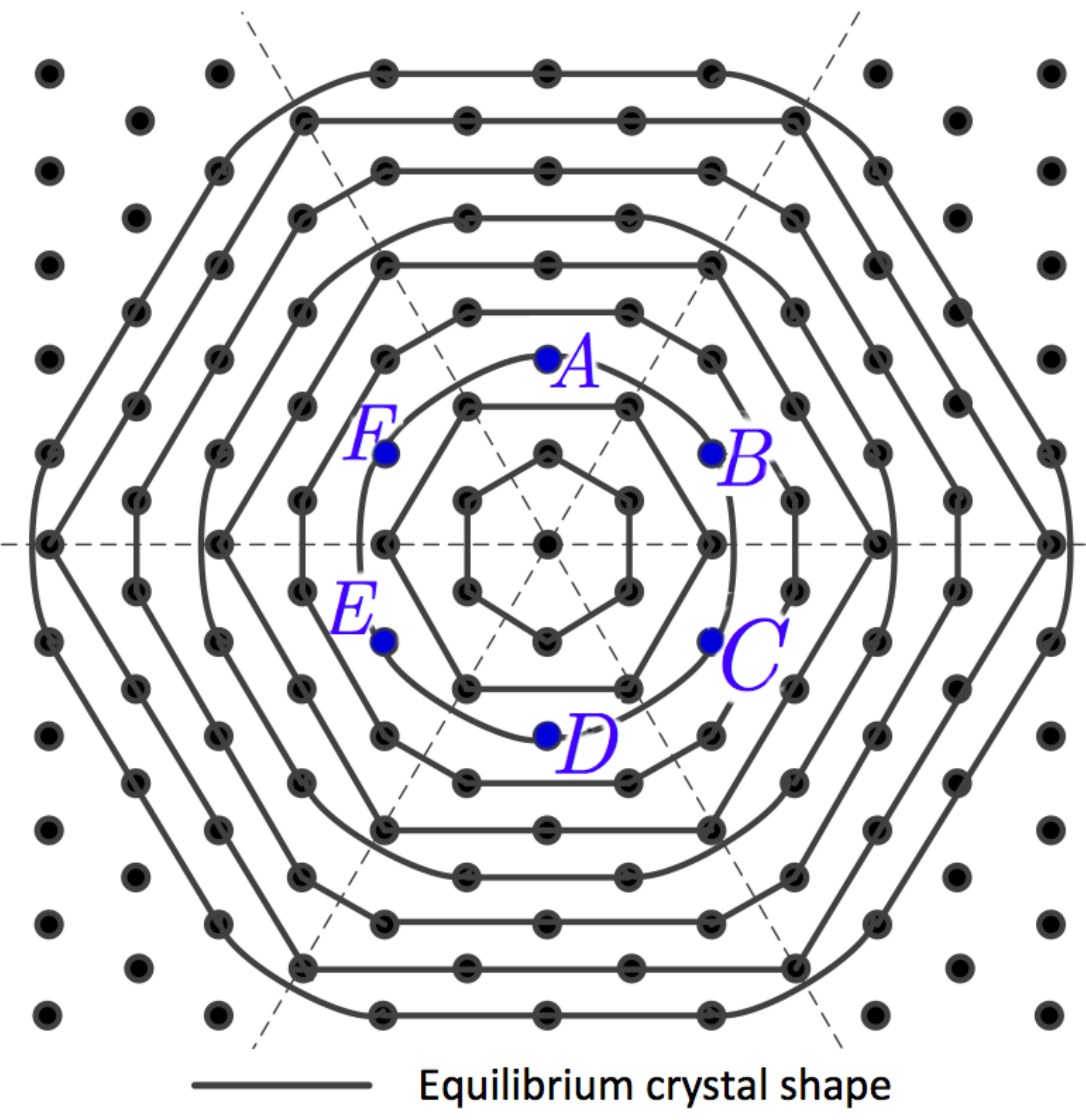}}
\caption{ 
  Equilibrium crystal shape used in the new geometric rule.
 }\label{Figure 10b}
\end{figure}

   From Figure \ref{Figure 10a}, we learn that the equilibrium crystal shape is a hexagon except for six narrow regions along the $30^\circ$-offset lattice axes where the transition from one edge of the hexagon to another edge is smoothed. The equilibrium crystal shape used in the new geometric rule is shown in Figure \ref{Figure 10b}.

   Figure \ref{Figure 10b} shows the equilibrium crystal shape used in the new geometric rule and the interface control neighbours of the cells. As an example, cells $A,B,C,D,E,F$ are on the same equilibrium crystal shape. Cells F,B are the interface control neighbours of $A$, cells $A,C$ are the interface control neighbours of $B$, etc.

The new geometric rule is applied after  Eq.~(\ref{(4)}): a new variable  $\delta_t (z)$ is defined to represent the {\it amount of water to be redistributed for cell $z$ at time $t$}, with initial value $\delta_t (z)=0$ for all $z$. 

\begin{defn}For a given cell $z_0$, define two {\rm interface control neighbours} $z_0^1,z_0^2$, which are two neighbouring cells of  $z_0$ on the same equilibrium crystal shape. 
   \end{defn}
   
   Define  $\overline s(z_0 )$ as the {\it average of the water} amounts in cell $z_0$ and its two interface control neighbours $z_0^1,z_0^2$, this is:
\begin{eqnarray}
\overline s(z_0 ):=\frac{1}{3} \left( \begin{array}{c}~\\ ~\end{array}\hspace{-0.1 in}s_{t+1}^- (z_0 )+s_{t+1}^- (z_0^1 )+s_{t+1}^- (z_0^2 )\right).
\end{eqnarray}

For every boundary $z_0$, if neither of  $z_0^1,z_0^2$ are frozen, then adjust  $\delta_t (z_0 )$ as follows
\begin{eqnarray}
\delta_s (z_0 )&=&\delta_s (z_0 )+\varepsilon(\overline s(z_0 )-s_{t+1}^- (z_0 )),\label{(13)}\\
\delta_s (z_0^1 )&=&\delta_s (z_0^1 )+\varepsilon(\overline s(z_0 )-s_{t+1}^- (z_0^1 )),\\
\delta_s (z_0^2 )&=&\delta_s (z_0^2 )+\varepsilon(\overline s(z_0 )-s_{t+1}^- (z_0^2 ))\label{(15)},
\end{eqnarray}
where $\varepsilon\in \mathbb{R}_{\geq 0}$ determines the amount of interface control. After $\delta_s (z)$ has been adjusted for all $z$ according to Eq.~(\ref{(13)})-(\ref{(15)}),   for every cell $z$ set 
\begin{eqnarray}
s_{t+1}^+ (z):=s_{t+1}^- (z)+\delta_s (z).\label{(16)}
\end{eqnarray}

 Recall that in the original Reiter's model, once water is accumulated in a boundary cell, water stays permanently in that cell. The new function Eq.~(\ref{(16)}) forces water redistribution particularly among boundary cells to smoothen the snow vapour interface. Figure \ref{Figure 11} below shows two snowflake images generated by the enhanced Reiter's model with the new geometric rule.

 \begin{figure}[h]
\centering
\resizebox{0.4\textwidth}{!}{%
  \includegraphics{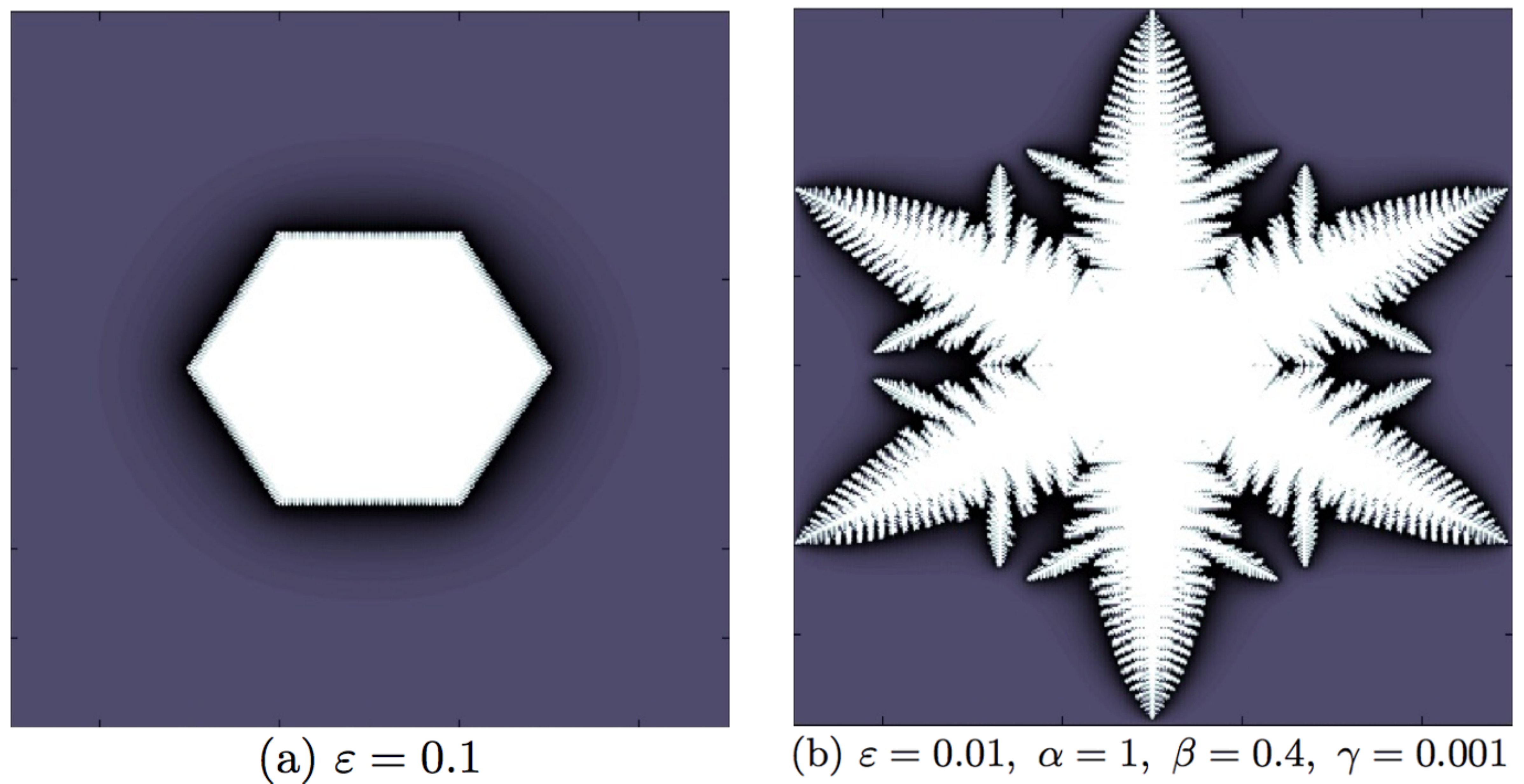} }
\caption{ 
Snowflake images generated by the enhanced Reiter's model with the new geometric rule. 
 }\label{Figure 11}
\end{figure}

At $\varepsilon=0.1$, the image above resembles a plate observed in nature much more closely than the ones in Figure \ref{Figure 2}. By reducing interface control with $\varepsilon=0.01$, the snowflake starts as a plate and later becomes a dendrite as diffusion control dominates interface control. 

 \section{Conclusions and future work \label{future}
}

In this paper we have analyzed the growth of snowflake images generated by a computer simulation model (Reiter's model \cite{[11]}), and have proposed ways to improve the model.  We have derived an analytical solution of the main branch growth latency and made numerical comparison with simulation results. Subsequently we observed interesting patterns of side branches in terms of growth latency and direction. Finally, to enhance the model, we have introduced a new geometric rule that incorporates interface control, a basic mechanism of the solidification process, which is not present in the original Reiter's model.

The present work has shed light into some   interesting patterns that lead to further questions about crystal growth. On the main branch growth, one may ask why the growth latency is almost constant (Figure \ref{Figure 5}) and whether this phenomenon is unique to the hexagonal cells or applicable to other two dimensional lattices. 
Concerning the  side branch growth, it was noted that some side branches grow much faster than their neighbours, and that with slightly different diffusion parameters the side branch growth latency could change drastically at the same position while the main branch growth latency remains virtually the same. The  study in Section \ref{enhanced} shows that this great sensitivity is attributable to diffusion instability - when the growth of cells in some direction gain initial advantage over their neighbours, the advantage continues to expand such that the growth in that direction becomes even faster. It was noted in Section \ref{enhanced}  that diffusion instability is caused by competition among cells in diffusion,  and  thus the average number of contributing neighbours is a good indicator to explain diffusion instability.
  Finally,    the enhanced model described in Section \ref{enhanced} can be used to explore the interplay of diffusion and interface control. For example, one may simulate growth in an environment where the diffusion and interface control parameters vary with time so as to generate images similar to Figure \ref{Figure 1} (b)(c).
 
 Recently Reiter's model was   used in the study of snowfall retrieval algorithms (e.g. see \cite{algorithms, falling} and references therein), and it was suggested that other mechanisms of snowflake formation from ice crystals besides aggregation must be considered in snowfall retrieval algorithms. It is thus natural to ask whether the enhanced Reiter's model constructed here may provide novel insights in this direction, as well as when considering crystal growth dynamics as in \cite{Kenneth}. Moreover, since cellular automata models have been considered for numerical computations of pattern formation in snow crystal growth, it would be interesting analyze the outcome of the implementation of the model presented here to the analysis done  in  \cite{pattern, Kelly}.  
 
 Finally,   the effects of  lattice anisotropy coupled to a diffusion process have been studied in \cite{Nigel} to understand phase diagrams associated to crystal growth.  Since this approach seemed recently useful from different perspectives (e.g. see \cite{Liu} and references therein), it would be interesting to study the   enhanced model constructed here from the perspective of \cite{Nigel, Liu}.  
 

 
 \noindent {\bf Acknowledgments:} The research in this paper was conducted by the first author under the supervision of the latter, as part of the IGL $\&$ MIT-PRIMES program. Both authors are thankful for the opportunity to conduct this research together, and would also  like to thank K. Libbrecht and N. Goldenfeld for helpful comments.




 
%



 \begin{small}
  
 \end{small}

\end{document}